\documentclass{ws-p8-50x6-00}

\usepackage{graphicx}
\usepackage{amsmath}
\usepackage{latexsym}

\newcommand{\lsim}{\mathrel{\rlap{\lower4pt\hbox{\hskip0pt$\sim$}}
\raise1pt\hbox{$<$}}}           
\newcommand{\gsim}{\mathrel{\rlap{\lower4pt\hbox{\hskip0pt$\sim$}}
\raise1pt\hbox{$>$}}}           

\newcommand{\sfrac}[2]{\mbox{\footnotesize $\frac{#1}{#2}$}}

\begin{document}

\title{The Character of Goldstone Bosons}

\author{M.B. Hecht,\footnotemark[1]$\,$ C.D. Roberts\footnotemark[1]$\,$ and
S.M. Schmidt\footnotemark[2]}

\address{\footnotemark[1]$\,$Physics Division, Argonne National Laboratory,\\
Argonne IL 60439, USA\\[0.5ex]
\footnotemark[2]$\,$Institut f\"ur Theoretische Physik, Universit\"at
T\"ubingen, \\D-72076 T\"ubingen, Germany}


\maketitle

\abstracts{A succinct review of the QCD gap equation and dynamical chiral
symmetry breaking; their connection with Bethe-Salpeter equations and resolving
the dichotomous nature of the pion; the calculation of the pion's valence-quark
distribution; and first results for the $\pi$-exchange contribution to the
$\gamma N \to \omega N$ cross-section, which is important in the search for
missing nucleon resonances.\vspace*{-1ex}}

\section{Introduction}\label{sec:Intro}
In the strong interaction spectrum the pion is identified as both a Goldstone
mode, associated with Dynamical Chiral Symmetry Breaking (DCSB), and a bound
state composed of $u$- and $d$-quarks.  This dichotomy is remarkable because,
while $m_\rho/2 \sim m_N/3 \sim 350\,$MeV$ =: M_q$, the constituent quark mass,
$m_\pi/2$ is only $\approx 0.2 M_q$; i.e., the pion is much lighter than
comparable bound states.  $M_q$ is the ``effective mass'' of quarks in a
hadron; and the ratio of this to the renormalisation-group-invariant $u$- and
$d$-current-quark masses ($\hat m \sim 10\,$MeV) indicates the magnitude of the
effect of nonperturbative dressing on light-quark propagation characteristics.
The peculiar nature of the pion can be expressed by the question: ``How does
one form an almost-massless bound state from very massive constituents {\em
without} fine-tuning?'' Answering this question; i.e., resolving and
understanding the dichotomy of the pion, is a key to unravelling the
quark-gluon substructure of QCD's Goldstone mode.

\section{QCD's Gap Equation}
The insightful study of dynamical symmetry breaking can be pursued using a {\em
gap equation}, which in QCD is the Dyson-Schwinger equation
(DSE)\cite{revbasti,reinhardreview} for the dressed-quark propagator:
\begin{equation}
\label{gendse} S(p)^{-1} = Z_2 \,(i\gamma\cdot p + m_{\rm bare}) +\, Z_1
\int^\Lambda_q \, g^2 D_{\mu\nu}(p-q) \frac{\lambda^a}{2}\gamma_\mu S(q)
\Gamma^a_\nu(q,p) \,.
\end{equation}
In this equation: $D_{\mu\nu}(k)$ is the renormalised dressed-gluon propagator;
$\Gamma^a_\nu(q;p)$ is the renormalised dressed-quark-gluon vertex; $m_{\rm
bare}$ is the $\Lambda$-dependent current-quark bare mass that appears in the
Lagrangian; and $\int^\Lambda_q := \int^\Lambda d^4 q/(2\pi)^4$ represents
mnemonically a {\em translationally-invariant} regularisation of the integral,
with $\Lambda$ the regularisation mass-scale.
In Eq.~(\ref{gendse}), $Z_1(\zeta^2,\Lambda^2)$ and $Z_2(\zeta^2,\Lambda^2)$
are the quark-gluon-vertex and quark wave function renormalisation constants,
which depend on the renormalisation point, $\zeta$, and the regularisation
mass-scale, as does the mass renormalisation constant
\begin{equation}
\label{Zmass} Z_m(\zeta^2,\Lambda^2) =
Z_4(\zeta^2,\Lambda^2)/Z_2(\zeta^2,\Lambda^2) ,
\end{equation}
with the renormalised mass given by $ m(\zeta) := m_{\rm
bare}(\Lambda)/Z_m(\zeta^2,\Lambda^2)$.

The solution of Eq.~(\ref{gendse}) has the form
\begin{equation}
\label{sinvp} S(p)^{-1} = i \gamma\cdot p \,A(p^2,\zeta^2) + B(p^2,\zeta^2)
        = \frac{1}{Z(p^2,\zeta^2)}\left[ i\gamma\cdot p +
        M(p^2,\zeta^2)\right]\,,
\end{equation}
where the functions $A(p^2,\zeta^2)$, $B(p^2,\zeta^2)$ express the effects of
quark-dressing induced by the quark's interaction with its own gluon field.
Equation~(\ref{gendse}) must be solved subject to a renormalisation [boundary]
condition, and in QCD it is practical to impose the requirement that at a large
spacelike $\zeta^2$
\begin{equation}
\label{renormS} \left.S(p)^{-1}\right|_{p^2=\zeta^2} = i\gamma\cdot p +
m(\zeta)\,.
\end{equation}

The dressed-quark mass function:
$M(p^2,\zeta^2)=B(p^2,\zeta^2)/A(p^2,\zeta^2)$, is actually independent of the
renormalisation point; i.e.,
it is a function only of $p^2/\Lambda_{\rm QCD}^2$.  At one loop order in
perturbation theory the running mass
\begin{equation}
\label{masanom} m(\zeta) = M(\zeta^2) = \frac{\hat m}
{\left(\rule{0mm}{1.2\baselineskip} \sfrac{1}{2}\ln\left[\zeta^2/\Lambda_{\rm
QCD}^2 \right]\right)^{\gamma_m}}\,,\; \gamma_m= 12/(33-2 N_f)\,,
\end{equation}
where $\hat m$ is the renormalisation point independent current-quark mass, and
the mass renormalisation constant in Eq.~(\ref{Zmass}) is
$Z_m(\zeta^2,\Lambda^2) = \left[
\alpha(\Lambda^2)/\alpha(\zeta^2)\right]^{\gamma_m}$.  (At one-loop in
Landau-gauge, $Z_2\equiv 1$.)

The chiral limit is unambiguously defined by\cite{mrt98} $\hat m = 0$.  In this
case there is no perturbative contribution to the scalar piece of the quark
self energy; i.e., $B(p^2,\zeta^2)\equiv 0$ at every order in perturbation
theory.

\section{Dynamical Chiral Symmetry Breaking}\label{sec:DCSB}
\parbox{33.5em}{
DCSB is the appearance of a $B(p^2,\zeta^2)\not\equiv 0$ solution of
Eq.~(\ref{gendse}) in the chiral limit, which is impossible in perturbation
theory.  Hence here, as in all studies of dynamical symmetry breaking, a
nonperturbative analysis of the gap equation is required. For that one needs a
systematic, symmetry-preserving truncation scheme, which goes beyond the weak
coupling expansion that yields perturbation theory.
}

The leading term in one such scheme\cite{truncscheme} is the
renormalisation-group-improved rainbow approximation to the gap
equation:\cite{mr97}
\begin{equation}
\label{ouransatz}
 S(p)^{-1} = Z_2 \,(i\gamma\cdot p + m_{\rm bare}) +
 \int^\Lambda_q \, {\cal G}((p-q)^2)\, D_{\mu\nu}^{\rm
free}(p-q) \frac{\lambda^a}{2}\gamma_\mu S(q) \frac{\lambda^a}{2}\gamma_\nu \,,
\end{equation}
where $D_{\mu\nu}^{\rm free}(k)$ is the free gauge boson propagator and ${\cal
G}(k^2)$ is an ``effective coupling''.  In QCD ${\cal G}(k^2)=4\pi \alpha(k^2)$
for $k^2\gsim 1\,$GeV$^2$, where $\alpha(k^2)$ is the strong running coupling
constant.  However, the behaviour of ${\cal G}(k^2)$ for $k^2< 1\,$GeV$^2$;
i.e., at infrared length-scales ($\gsim 0.2\,$fm), is currently unknown.

The effective coupling is a phenomenological mnemonic for the contracted
product of the dressed-gluon propagator and dressed-quark-gluon vertex, which
forms the kernel in Eq.~(\ref{gendse}). The infrared strength of the
interaction can then be characterised by the
interaction-tension\cite{cdrvienna}
\begin{equation}
\label{IT}
 \sigma^{\Delta}:= \frac{1}{4\pi}\int_{\Lambda_{\rm
QCD}^2}^{\Lambda_{\rm
pQCD}^2}\,dk^2\,k^2\,\left[\Delta(k^2)-\Delta(\Lambda_{\rm pQCD}^2)\right]
\end{equation}
with  $\Delta(k^2)={\cal G}({k^2})/k^2$, and $\Lambda_{\rm pQCD} =
10\,\Lambda_{\rm QCD}$, the boundary above which perturbation theory is
unquestionably valid.

Studies show$\,$\cite{fredIR} that DCSB does not occur for
$\sigma^{\Delta}\lsim 0.5\,$GeV$^2\sim 9\Lambda_{\rm QCD}^2$, which means that
the existence of a dynamically generated, $B\not\equiv 0$ solution of
Eq.~(\ref{gendse}) is impossible without a significant infrared enhancement of
the effective interaction. Reproducing observable phenomena
requires$\,$\cite{cdrvienna,mt99} $\sigma^{\Delta}\gsim 4\,{\rm GeV}^2\sim
70\,\Lambda_{\rm QCD}^2$; i.e., a ten-fold enhancement over the critical value.
The origin of this enhancement, whether in the gluon or ghost vacuum vacuum
polarisation, or elsewhere, is currently unknown but is actively being sought
(see, e.g., Refs.~[\ref{continuumgluonR},\ref{latticegluonR}]). Nevertheless,
its characterisation via a one-parameter model yields an efficacious
description of light meson observables.\cite{marisvienna}

The dynamical generation of a dressed-quark mass in the chiral limit is
necessarily accompanied by the formation of a vacuum quark
condensate:\cite{mrt98}
\begin{equation}
\label{qbq0} \,-\,\langle \bar q q \rangle_\zeta^0 = Z_4(\zeta^2,\Lambda^2)\,
N_c {\rm tr}_{\rm D}\int^\Lambda_q\,
        S^{0}(q,\zeta)\,,
\end{equation}
where ${\rm tr}_D$ identifies a trace over Dirac indices only and the
superscript ``$0$'' indicates the quantity was calculated in the chiral limit.
This condensate is gauge-parameter and cutoff independent.  In the model
summarised in Ref.~[\ref{marisviennaR}], an explanation of hadron observables
{\em requires} $\langle \bar q q \rangle_{\zeta=1\,{\rm GeV}}^0 = -(0.241\,{\rm
GeV})^3$.  The condensate is analogous to a Cooper pair density and this value
corresponds to $1.8\,$fm$^{-3}$.  Considering, simply for illustration, the
condensate to be a sea of close-packed spheres then this density corresponds to
a Cooper pair radius of $0.76\,$fm, which is just $15$\% more than the pion's
charge radius and $13$\% less than the nucleon's radius. Clearly the scale set
by this condensate is fundamental in QCD.

\section{The Goldstone Boson}
We began with a question: ``How does one form an almost-massless bound state
from very massive constituents {\em without} fine-tuning?''  Bound states of a
dressed-quark and dressed-antiquark; i.e., mesons, are described by the
homogeneous Bethe-Salpeter equation (BSE).  (The Bethe-Salpeter amplitude
obtained as the solution of this equation is analogous to the wave function in
quantum mechanics and provides similar insights into the nature of the bound
state.\cite{a1b1})  It was proven in Ref.~[\ref{mrt98R}] that the flavour
nonsinglet pseudoscalar meson BSE admits a massless solution if, and only if,
the QCD gap equation supports DCSB.  This is an expression of Goldstone's
theorem.

That proof establishes a number of corollaries.  One is a collection of
quark-level Goldberger-Treiman relations, which establish that the pseudoscalar
meson Bethe-Salpeter amplitude necessarily has pseudovector components.  It is
these components that are responsible for the asymptotic $1/Q^2$ behaviour of
the electromagnetic pion form factor.\cite{mrpion} Another is a mass formula,
valid for all flavour nonsinglet pseudoscalar mesons, {\em independent} of the
current-quark mass of the constituents.\cite{mishaSVY}  This single mass
formula unifies aspects of QCD's light- and heavy-quark sectors, and
provides\cite{cdrqciv} a qualitative understanding of recent results from
lattice simulations for the current-quark-mass-dependence of pseudoscalar meson
masses.\cite{cmichael} DCSB and the large value of the vacuum quark condensate
also explain the large $\pi$-$\rho$ mass difference.  An heuristic
demonstration of this is provided by the bosonisation procedure described in
Ref.~[\ref{justinR}].  In the chiral limit, for the flavour nonsinglet mesons,
there is an exact cancellation between the condensate-driven mass-term in the
quark-loop piece of the effective action and that in the auxiliary field piece,
while for the vector mesons, because $\{\gamma_5,\gamma_\mu\}=0$, the terms
add.

Confinement is the failure to observe coloured excitations in a detector.  It
is related to the effective interaction and interaction-tension described in
Sec.~\ref{sec:DCSB}.  If the interaction-tension is large enough to explain
hadron observables then the analytic properties of the dressed-quark propagator
are very different from those of a free fermion; e.g., the propagator no longer
has a K\"all\'en-Lehmann representation.  The absence of such a representation
is a sufficient condition for confinement.\cite{revbasti,reinhardreview}  From
this perspective, the existence of flavour nonsinglet pseudoscalar Goldstone
modes is a consequence of confinement: one can have Goldstone modes without
confinement, but one cannot have confinement in the chirally symmetric theory
without the appearance of Goldstone modes.

\begin{table}[t]
\caption{\label{tablea}Fit parameters for the calculated $u$ valence-quark
distribution function, Eq.~(\protect\ref{xuvx}).}
\begin{center}
\begin{tabular}{c|ccccc}
scale (GeV) & $A_u$ & $\eta_1$ & $\eta_2$ & $\epsilon_u $ & $\gamma_u$\\\hline
0.54 & 11.24 & 1.43 & 1.90 & 2.44 & 2.54\\
2.0~ & ~4.25 & 0.97 & 2.43 & 1.82 & 2.46 \\\hline
\end{tabular}
\end{center}
\end{table}

\section{Pion's Valence-quark Distribution}
The quark-gluon substructure of the pion is also expressed in its parton
distribution functions.  They can be measured in $\pi N$ Drell-Yan\cite{DYpion}
and deep inelastic scattering\cite{roy} (using the Sullivan process) but cannot
be calculated in perturbation theory.  The valence-quark distribution function
has been calculated\cite{hecht} using the DSE model that unified the small- and
large-$Q^2$ behaviour of the electromagnetic pion form factor.\cite{mrpion}
This covariant calculation of the relevant ``handbag diagrams'' indicates that,
at a resolving scale of $q_0=0.54\,$GeV$=1/(0.37\,{\rm fm})$, valence-quarks
with an active mass of $300\,$MeV carry $71$\% of the pion's momentum, and
yields a numerical result for the distribution function that is pointwise
well-described by
\begin{equation}
\label{xuvx}
 x\,u_V^\pi(x;q_0) = A_u\,x^{\eta_1}\,(1-x)^{\eta_2}\, ( 1 -
\epsilon_u\sqrt{x} + \gamma_u x)\,,
\end{equation}
with the parameter values listed in Table~\ref{tablea}.  This parametrisation,
with the parameter values listed, provides an equally good fit to $x\,
u_V^\pi(x;2\,{\rm GeV})$ obtained by applying the three-flavour, leading-order,
nonsinglet renormalisation group evolution equations.  The low moments of the
distribution calculated at this scale ($2\,$GeV) are$\,$\cite{hecht}
\begin{equation}
\label{tableb}
\begin{array}{l|lll}
 & \langle x \rangle & \langle x^2 \rangle & \langle x^3 \rangle \\ \hline
{\rm Calc.} & 0.24 & 0.098 & 0.049 \\
{\rm Exp.} & 0.24 \pm 0.01 & 0.10\pm 0.01 & 0.058\pm 0.004\\
{\rm Latt.} & 0.27 \pm 0.01  & 0.11 \pm 0.3 & 0.048 \pm 0.020\\\hline
\end{array}
\end{equation}
with the experimental results from Ref.~[\ref{DYpionR}] and the lattice results
from Ref.~[\ref{latticeR}].

The value of $\eta_2\simeq 2$ obtained in the DSE calculation is consistent
with a perturbative QCD analysis:\cite{matthias} $u_V^\pi(x)\sim (1-x)^2$ at
$x\simeq 1$. That prediction, however, conflicts with the currently available
Drell-Yan data,\cite{DYpion} which suggests that $u_V^\pi(x) \approx (1-x)$ for
$x\simeq 1$.  This experimental result, if true, will be a profound challenge
to QCD, even questioning whether the strong interaction Lagrangian contains a
vector interaction.

We note that the agreement between the moments calculated using the DSE result
and those inferred from experiment, see Eq.~(\ref{tableb}), indicates plainly
that the low moments are insensitive to the large-$x$ behaviour of the
distribution; i.e., the valence region: even the {\em sixth} moments of these
differently shaped distributions disagree by only $32$\%. Hence pointwise
calculations of the distribution functions are necessary to probe the valence
region.

\section{\mbox{\boldmath $\omega$}-meson Photoproduction}
Constituent quark models predict many nucleon resonances that are hitherto
unobserved.  That may be because they only couple weakly to the $\pi N$
channel, which has been used to search for them.   Experiments at facilities
such as JLab are therefore seeking the ``missing resonances'' in
electromagnetic reactions.  Of these, $\omega$-photoproduction is particularly
useful because its nonresonant reaction mechanisms are thought to be well
understood.\cite{harrylee}  Hence a comparison between experimental data and
the cross-section calculated using these mechanisms alone provides a sound base
from which to search for nucleon resonance contributions.

\begin{figure}[t]
\begin{center}
\begin{picture}(144,120)(-72,-72)
\setlength{\unitlength}{0.75pt}
\thicklines
\put (-72,-72){\line(2,1){72}} \put (-66,-60){$N$}
\put (0,-36){\line(2,-1){72}} \put (66,-60){$N$}
\put (0,-36){\circle*{12}}
\multiput (0,-36)(0,9){6}{\line(0,1){6}} \put (9,-9){$\pi^\ast$}
\put (0,18){\circle*{12}}
\put (0,18){\line(2,1){72}}\put (66,42){$\omega$}
%
%
\multiput (0,18)(-20,10){4}{\line(-2,1){14}} \put(-72,42){$\gamma$}
\end{picture}\vspace*{-4ex}
\end{center}
\caption{\label{drawing} $\pi$-exchange contribution to the
$\omega$-photoproduction amplitude.}\vspace*{-2em}
\end{figure}
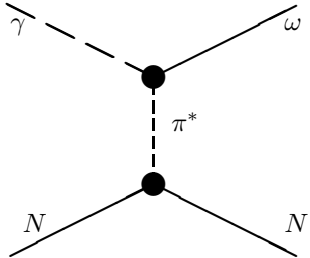

Meson exchange models\protect\cite{harrylee} suggest that the leading
contribution to the photoproduction cross-section at low energies and forward
angles is given by the off-shell-$\pi$-exchange diagram depicted in
Fig.~\ref{drawing}. Here the filled circles represent meson-nucleon and
meson-photon form factors, which are customarily parametrised in the
calculation of observables.

\begin{figure}[t]
\centerline{\includegraphics[height=35.0em]{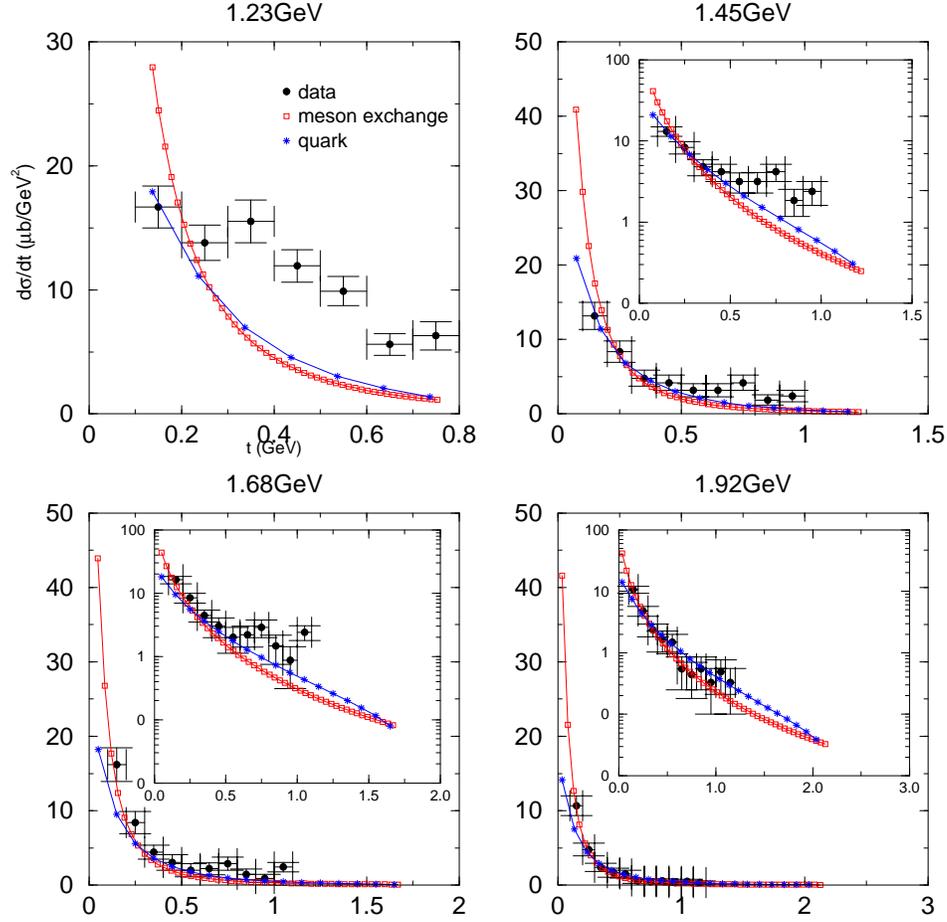}}
%
\caption{\label{figa} $\pi$-exchange contribution to the $\gamma N \to \omega
N$ cross-section obtained using the model of
Refs.~[\protect\ref{myriadR},\protect\ref{edmR}].  (Panels are labelled by the
incident photon energy.)  $\ast$, long-dashed line, our calculation (no
parameters were varied to obtain this result); $\Box$, short-dashed line, the
meson-exchange model calculation of Ref.~[\protect\ref{harryleeR}] (only the
$t$-channel $\pi$-exchange contribution is shown); data from
Ref.~[\protect\ref{saphirR}].}\vspace*{-2em}
\end{figure}

However, these form factors can be {\em calculated} when one has a reliable
quark-level description of hadrons.\cite{myriad} We illustrate that using the
model of Ref.~[\ref{myriadR}], with the {\it Ansatz} for the nucleon's Fadde'ev
amplitude updated as described in Ref.~[\ref{edmR}], to obtain a {\em
parameter-free} prediction of the contribution from the $\pi$-exchange process
to the $\omega$-photoproduction cross-section.  The results are illustrated in
Fig.~\ref{figa}.  Much of the discrepancy at large angles owes to our neglect,
in this preliminary calculation, of the contribution from $s$- and $u$-channel
nucleon diagrams. Including them is part of an ongoing programme.

\section{Epilogue}
The DSEs are maturing as an efficacious nonperturbative tool in hadron physics
and in this application they provide crucial information about the long-range
part of the QCD interaction.\cite{a1b1}  The primary elements in the approach
are the Schwinger functions: the dressed-propagators and -vertices.  They can
also be calculated using other techniques, such as effective field theory and
lattice quantisation methods.  There is therefore great scope for forging
relationships between these approaches, thereby providing a tool whose domain
of reliable application is larger than that of its individual parts.  One
concrete example is the confirmation in recent lattice-QCD
simulations\cite{latticequark} of the bulk features of the dressed-quark
propagator that have long been suggested by DSE
studies.\cite{marisvienna,cdrqciv} Making that agreement quantitative will hone
our understanding of the long-range part of the interaction and, e.g., make
possible a DSE-built bridge between lattice-QCD and the $Q^2$-evolution of the
electromagnetic pion form factor.\cite{Fpi}
\vspace*{-2ex}

\medskip

\section*{Acknowledgments}
%
%
CDR gratefully acknowledges the hospitality of the
staff at the Special Research Centre for the Subatomic Structure of Matter at
the University of Adelaide, and also financial support provided by the
Centre.
This work was also supported by: the Deutsche For\-schungs\-ge\-mein\-schaft
under project no.\ SCHM~1342/3-1; and the US Department of Energy, Nuclear
Physics Division, under contract no.\ W-31-109-ENG-38.

\end{document}